\documentclass[12pt]{article}
\usepackage{epsfig}
\title{ Investigation of the $e^+e^- \to \omega \pi^0 \to \pi^0 \pi^0 \gamma$
reaction in the energy domain near the $\phi$-meson}
\author
{ M.~N.~Achasov, V.~M.~Aulchenko, A.~V.~Berdyugin, A.~V.~Bozhenok, \\
D.~A.~Bukin, S.~V.~Burdin, T.~V.~Dimova, V.~P.~Druzhinin, \\
M.~S.~Dubrovin, I.~A.~Gaponenko, V.~B.~Golubev, V.~N.~Ivanchenko, \\
I.~A.~Koop, A.~A.~Korol, S.~V.~Koshuba, E.~V.~Pakhtusova, \\
A.~A.~Salnikov, V.~V.~Shary, S.~I.~Serednyakov, Yu.~M.~Shatunov, \\
V.~A.~Sidorov, 
Z.~K.~Silagadze \thanks {Corresponding author. Fax +7 3832 34 21 63,
e-mail silagadze@inp.nsk.su} ~, A.~A.~Valishev
\vspace*{3mm} \\ Budker Institute of Nuclear Physics,  630 090,
Novosibirsk, Russia \\ \vspace*{1mm}
Novosibirsk State University, 630 090, Novosibirsk, Russia  }
\date{}

\begin{document}
\large
\maketitle

\begin{abstract}
The $e^+e^- \to \omega \pi^0 \to \pi^0 \pi^0 \gamma$ process was investigated
in the SND experiment at the VEPP-2M collider. A narrow energy interval near
the $\phi$-meson was scanned. The observed cross-section reveals, at the level
of three standard deviation, the interference effect caused by $\phi \to
\pi^0\pi^0\gamma$ decay. The cross-section parameters, as well as the real
and imaginary parts of the $\phi$-meson related amplitude, were measured.
\end{abstract}

\section{Introduction}
There are several reasons \cite{ND,NDR}, which make the 
experimental study of the $e^+e^-\to\omega\pi^0$ reaction interesting. 
First of all, it is expected that the radial excitations 
of $\rho$-meson should
reveal themselves in this process, thus giving us a possibility to study
their properties.  
The $e^+e^-\to\omega\pi^0$ transition is also important for the precise 
determination of the total cross section of $e^+e^-$ annihilation into 
hadrons. 
In the vector meson dominance model $e^+e^-\to\omega\pi^0$ transition is
connected to the $\omega\rho\pi$ vertex, which appears also in a number of
hadron decays, like $\omega\to 3\pi$, $\omega\to\pi^0\gamma$, 
$\rho\to\pi\gamma$, $\omega\to\mu^+\mu^-\pi^0$, $\pi^0\to 2\gamma$. The
precise experimental data about these processes stimulate theoretical
study of the underlying hadron dynamics. In the low energy limit, the chiral
perturbation theory \cite{ChPT} and effective chiral Lagrangians \cite{ChLag}
give a phenomenological description of meson physics, as it is
commonly believed nowadays. But their applicability is in question for 
energies above 1 GeV. On the other hand, for energies $1\div2$ GeV, the
perturbative QCD is also not applicable. So any precise experimental 
information in this energy range can be considered as "data in searching
of the theory". Simple but successful vector meson dominance picture may
become insufficient when the precision of the data increases.

The $e^+e^-\to\omega\pi^0$ transition can be studied in either 
$e^+e^-\to\omega
\pi^0\to\pi^+\pi^-\pi^0\pi^0$ or $e^+e^-\to\omega\pi^0\to\pi^0\pi^0\gamma$
channels. The former can provide about one order of magnitude more statistics,
but the latter is more preferable concerning background conditions.

The $e^+e^-\to\omega\pi^0\to\pi^0\pi^0\gamma$ reaction was studied earlier 
in the energy range
$1.0\div 1.4$ GeV by the ND detector \cite{ND,NDR}. 
Indirectly,
the $\sigma(e^+e^-\to\omega\pi^0)$ cross section was extracted also from
the ARGUS data on $\tau^-\to\nu_\tau\omega\pi^-$ decay \cite{ARGUS}, 
under assumption of the Conserved Vector Current (CVC). 
The results are in good agreement, thus confirming the 
CVC hypothesis \cite{CVC}.

In 1995 a new set of experiments began with the SND detector 
\cite{SND,SNDPR,BEGEXP} on the Novosibirsk
VEPP-2M storage ring \cite{VEPP2M}. 
The maximum
luminosity of VEPP-2M at $2E=1{\rm GeV}$ equals $ 3\cdot 10^{30} cm^{-2}
sec^{-1}$. 

Below we report the results on $e^+e^-\to\omega\pi^0\to\pi^0\pi^0\gamma$
process, based on 1996-1997 two-year SND statistics.
In 1996 several scans were made of the energy interval $2E=984 \div 1040 
{\mathrm MeV}$
\cite{EXP96}. From the 1997 experiment \cite{EXP97} a part of statistics,
collected at the center of mass energies 980, 1040 and 1060 MeV, was used.

\section{Theoretical Model}
As we have already mentioned in the introduction, currently no definite
predictions can be made from the basic 
first-principle theory of strong interactions (QCD) in this energy range.
The limited accuracy of the existing experimental data leaves enough room
for various phenomenologically inspired 
models. Now a light is seen at the end of the tunnel: several meson
factories come into operation, and also a new generation of fixed-target
experiments will accumulate a huge number of events with strongly 
interacting particles. So the experimental information is expected 
to become increasingly precise. 
Of course a common wisdom says that ``If we see
a light at the end of the tunnel, it's the light of an oncoming train''
\cite{Jam}.
So it is not excluded that the present theoretical models will be shattered
by this ``train'', but it can deliver a new passenger also.

Here we consider a simple phenomenological model, based on the vector meson
dominance scheme. Our aim is twofold: to estimate the expected 
cross section and to establish a framework for Monte-Carlo simulation.

\subsection{General considerations}
Let $J_\mu$ be the matrix element of the electromagnetic current between
the vacuum and the $\pi^0 \pi^0 \gamma$ final state. Then the amplitude
for the  $e^+e^- \to \pi^0 \pi^0 \gamma$ transition via one-photon 
annihilation diagram

\input FEYNMAN
\begin{center}
\begin{picture}(20000,15000)
\drawline\fermion[\SE\REG](550,13000)[8000]
\drawarrow[\NW\ATBASE](\pmidx,\pmidy)
\global\advance\pfrontx by -2000
\put(\pfrontx,\pfronty){$p_+$}
\drawline\fermion[\SW\REG](\pbackx,\pbacky)[8000]
\drawarrow[\NE\ATBASE](\pmidx,\pmidy)
\global\advance\pbackx by -2000
\put(\pbackx,\pbacky){$p_-$}
\drawline\photon[\E\REG](\pfrontx,\pfronty)[6]
\put(\photonbackx,\photonbacky){\circle*{2000}}
\drawline\photon[\SE\REG](\photonbackx,\photonbacky)[9]
\global\advance\pbackx by 650
\put(\pbackx,\pbacky){$k$}
\drawline\scalar[\NE\REG](\photonfrontx,\photonfronty)[4]
\global\advance\pbackx by 650
\put(\pbackx,\pbacky){$q_1$}
\drawline\scalar[\E\REG](\photonfrontx,\photonfronty)[3]
\global\advance\pbackx by 650
\put(\pbackx,\pbacky){$q_2$}
\end{picture}
\end{center}

\noindent reads (up to irrelevant phase factor)
$$M(e^+e^- \to \pi^0 \pi^0 \gamma)=\frac{e}{s}\bar v(p_+)\gamma_\mu
u(p_-)J^\mu \; , \; s=(p_++p_-)^2 .$$
\noindent The standard considerations lead to the following
cross section with unpolarized beams \cite{Jobra}
\begin{eqnarray}
d\sigma(e^+e^- \to \pi^0 \pi^0 \gamma)=\frac{\pi \alpha}{s^2}
\left ( \overline{J_1^* \cdot J_1}+\overline{J_2^* \cdot J_2} 
\right ) d\Phi ,
\label{cs} \end{eqnarray} \noindent
where only transverse (with respect to the beam direction) components
of $\vec{J}$ contribute and the invariant phase space element is
$$d\Phi=\frac{d\vec{q_1}}{(2\pi)^32E_1}\frac{d\vec{q_2}}{(2\pi)^32E_2}
\frac{d\vec{k}}{(2\pi)^32\omega}(2\pi)^4\delta(Q-q_1-q_2-k).$$
$J_\mu$ should satisfy the current conservation condition $Q_\mu J^\mu=0$.
On the other hand $J_\mu=T_{\mu \nu}\epsilon^\nu$, $\epsilon^\nu$ being
the photon polarization 4-vector. So $T_{\mu \nu}$ is a gauge invariant
tensor:
$$Q_\mu T^{\mu \nu}=k_\nu T^{\mu \nu} = 0 .$$
There exists a general procedure how to construct such gauge invariant
tensors which are free from kinematical singularities \cite{Tung}. In
our case there are just three independent tensors \cite{Kuraev}
\begin{eqnarray} &&
L^{(1)}_{\mu \nu}=(k\cdot Q)g_{\mu \nu}-k_\mu Q_\nu , \;
Q=q_1+q_2+k=p_-+p_+, \; q=\frac{1}{2}(q_1-q_2) \nonumber \\ &&
L^{(2)}_{\mu \nu}=(k\cdot Q)q_\mu q_\nu -(k\cdot q)(k_\mu q_\nu+
q_\mu Q_\nu) +(k\cdot q)^2g_{\mu \nu}
\nonumber \\ &&
L^{(3)}_{\mu \nu}=(k\cdot q)[Q^2g_{\mu \nu}-Q_\mu Q_\nu] +
q_\nu [(k\cdot Q)Q_\mu-Q^2k_\mu]. \nonumber \end{eqnarray}
\noindent So
$$T_{\mu \nu}=\frac{A_1}{E^2}L^{(1)}_{\mu \nu}+
\frac{A_2}{E^4}L^{(2)}_{\mu \nu}+\frac{A_3}{E^4}L^{(3)}_{\mu \nu} \; .$$
\noindent dimensionless form factors $A_i$ are determined by the concrete
dynamical model. Summing over the final state photon polarizations
(REDUCE program \cite{reduce} proves to be useful for these calculations)
and performing the angular integration, we get \cite{Kuraev}
\begin{eqnarray} &&
\frac{1}{4}\left ( \overline{J_1^* \cdot J_1}+\overline{J_2^* \cdot J_2} 
\right )=-\frac{1}{4} (T_1^{\nu *}T_{1 \nu}+T_2^{\nu *}T_{2 \nu})
\longrightarrow F(x,x_1,x_2)=\nonumber \\ && 
C_{11}|A_1|^2+C_{22}|A_2|^2+C_{33}|A_3|^2+ 
C_{12}(A_1A_2^*+A_1^*A_2)+ \\ && C_{13}(A_1A_3^*+A_1^*A_3)+
C_{23}(A_2A_3^*+A_2^*A_3), \nonumber \label{evme} \end{eqnarray} \noindent
where ($x=\frac{\omega}{E}=2-x_1-x_2, x_i=\frac{E_i}{E}, r=\frac{m_\pi}
{E}$)
\begin{eqnarray} && 
C_{11}=\frac{4}{3}x^2 \; \; , \; C_{13}=\frac{8}{3}x(x_1-x_2) \; \; ,
\; C_{23}=\frac{4}{3}(x_1-x_2)^3 \; , \nonumber \\ && 
C_{22}=\frac{1}{3}(x_1-x_2)^4+\frac{2}{3}x^2 ( r^2- 
1+x )^2 +\frac{2}{3}(x_1-x_2)^2(1-x)( r^2-1+x ) 
\nonumber \\ && 
C_{33}=\frac{8}{3}(x_1-x_2)^2(1+x)-\frac{8}{3}x^2( r^2-
1+x ) \\ &&
C_{12}=\frac{2}{3}\left [(x_1-x_2)^2+x^2 (r^2-1+x)\right ].
\nonumber \label{cij} \end{eqnarray}
So for the total cross-section we obtain (the factor $\frac{1}{2}$ accounts
for identical $\pi^{0,}s$)
\begin{eqnarray} &&
\sigma(e^+e^-\to \pi^0 \pi^0 \gamma)=\frac{\alpha}{256\pi^2E^2}
\int \limits_{x_{1-}}^{x_{1+}}dx_1 \int \limits_{x_{2-}}^{x_{2+}}dx_2
~F(2-x_1-x_2,x_1,x_2)\nonumber \\ && =\frac{\alpha}{256\pi^2E^2}
\int \limits_{x_{-}}^{x_{+}}dx \int \limits_{x^*_{2-}}^{x^*_{2+}}dx_2
~F(x,2-x-x_2,x_2) \; . \label{sg} \end{eqnarray}
The integration limits are determined from the condition
$$ |\cos{\theta_{12}}|=\left | \frac{(2E-E_1-E_2)^2-\vec{q_1}^2-
\vec{q_2}^2}{2|\vec{q_1}||\vec{q_2}|} \right | \le 1 \; , $$
\noindent which gives
\begin{eqnarray} &&
x_{1-}=r \; , \; x_{1+}=1 \; , \; x_-=0 \; , \; x_+=1-
r^2 \; , \nonumber \\ &&
x_{2\pm}=\frac{1}{1-x_1+\frac{r^2}{4}} \left \{ \left (1-\frac{x_1}{2}
\right ) \left ( 1-x_1+\frac{r^2}{2} \right ) \right . \pm\nonumber\\ &&
\left . \frac{1}{2} (1-x_1) \sqrt {x_1^2-r^2 } \right \}\; \; , 
\; \; x^*_{2\pm}=1+\frac{x}{2} \left [-1\pm \sqrt { \frac {1-
r^2-x} {1-x} } \right ]\; . \label{intlim} \end{eqnarray}

\subsection{Vector mesons contribution}
To proceed, we need expressions for the $A_i$ form factors. It is expected
that the main contribution comes from diagrams of the following type

\begin{picture}(30000,19000)
\drawline\fermion[\SE\REG](550,15000)[8000]
\drawarrow[\NW\ATBASE](\pmidx,\pmidy)
\drawline\fermion[\SW\REG](\pbackx,\pbacky)[8000]
\drawarrow[\NE\ATBASE](\pmidx,\pmidy)
\drawline\photon[\E\REG](\pfrontx,\pfronty)[6]
\put(\photonbackx,\photonbacky){\circle*{600}}
\drawline\fermion[\S\REG](\photonbackx,\photonbacky)[300]
\drawline\fermion[\E\REG](\pbackx,\pbacky)[6000]
\drawline\fermion[\N\REG](\photonbackx,\photonbacky)[300]
\drawline\fermion[\E\REG](\pbackx,\pbacky)[6000]
\global\advance\pmidy by 1000
\put(\pmidx,\pmidy){$V_1$}
\drawline\fermion[\S\REG](\pbackx,\pbacky)[600]
\put(\pmidx,\pmidy){\circle*{600}}
\drawline\scalar[\SE\REG](\pmidx,\pmidy)[5]
\global\advance\pbackx by 1000
\put(\pbackx,\pbacky){$q_1$}
\global\advance\pbackx by 1000
\drawline\fermion[\NW\REG](\pfrontx,\pfronty)[300]
\drawline\fermion[\NE\REG](\pbackx,\pbacky)[4000]
\drawline\fermion[\SE\REG](\pbackx,\pbacky)[600]
\put(\pmidx,\pmidy){\circle*{600}}
\drawline\fermion[\SW\REG](\pbackx,\pbacky)[4000]
\global\advance\pmidx by -2000
\global\advance\pmidy by 700
\put(\pmidx,\pmidy){$V_2$}
\drawline\scalar[\SE\REG](\pfrontx,\pfronty)[3]
\global\advance\pbackx by 1000
\put(\pbackx,\pbacky){$q_2$}
\global\advance\pbackx by 4000
\put(\pbackx,\pbacky){$+$}
\global\advance\pbackx by 3000
\put(\pbackx,\pbacky){$(q_1\longleftrightarrow q_2)$}
\global\advance\pfronty by 400
\drawline\photon[\NE\REG](\pfrontx,\pfronty)[7]
\global\advance\photonbackx by 1000
\put(\photonbackx,\photonbacky){$k$}
\end{picture}

Using kinematical structures of the $V_1 \to V_2\pi^0$ and $V\to \pi^0
\gamma$ vertexes, which are determined from the Lorentz covariance, we
get for the form factors \cite{Kuraev}
\begin{eqnarray} &&
A_1=-\frac{1}{4} \left (\frac{m^2}{E^2}-1+\frac{3}{2}x \right )
[g(x_1)+g(x_2)]+\frac{1}{4}(x_1-x_2)[g(x_1)-g(x_2)], \nonumber \\ &&
A_2=\frac{1}{4}[g(x_1)+g(x_2)] \; \; , \; \; 
A_3=-\frac{1}{8}[g(x_1)-g(x_2)] \; , \label{Ai} \end{eqnarray}
\noindent where
$$g(x)=\sum \limits_{V_1,V_2=\rho,\omega,\phi} g^{(V_1,V_2)}(x)\; , $$
\noindent and individual contributions look like (note that $V\to\pi\gamma$
coupling constant is defined as $eg_{V\pi\gamma}$)
\begin{eqnarray} &&
 g^{(V_1,V_2)}(x)=\pi\alpha M_{V_1}^2~\frac{g_{V_1V_2\pi}g_{V_2\pi\gamma}}
{g_{V_1}}~\frac{s}{s-M_{V_1}^2+iM_{V_1}\Gamma_{V_1}}
~\frac{1}{1-x+\gamma_{V_2}} \; ,
\nonumber \\ &&
\gamma_V=\frac{1}{s}(m^2-M_V^2+iM_V\Gamma_V). \label{gx} \end{eqnarray}
To take into account the resonance width dependence on energy, one should
replace $M_V\Gamma_V \to \sqrt{q^2}\Gamma_V(q^2)$, $q$ being the resonance 
4-momentum.

\subsection{Coupling constants}
$g_{V\pi\gamma}$ can be determined from the $V\to\pi^0\gamma$ decay width.
Namely
$$g^2_{V\pi\gamma}=\frac{24}{\alpha}\frac{M_V^3\Gamma(V\to\pi^0\gamma)}
{(M_V^2-m_\pi^2)^3} \; .$$
Using $\Gamma(\omega\to\pi^0\gamma)$, $\Gamma(\rho^+\to\pi^+\gamma)$, and
$\Gamma(\phi\to\pi^0\gamma)$ as inputs, we get
\begin{eqnarray} &&
g_{\omega\pi\gamma}=(2.32\pm 0.06){\rm GeV}^{-1} \; , \;
g_{\rho\pi\gamma}=(0.73\pm 0.04){\rm GeV}^{-1} \; , \nonumber \\ &&
g_{\phi\pi\gamma}=(0.138\pm 0.007){\rm GeV}^{-1} \; . \label{gwpg}
\end{eqnarray}
As for the photon--vector meson coupling constant, it can be determined from
the $\Gamma(V\to e^+e^-)$ width
$$\frac{g_V^2}{4\pi}=\frac{\alpha^2}{3}\frac{M_V}{\Gamma(V\to e^+e^-)}\; ,$$
\noindent and we get
\begin{eqnarray}
g_\rho=5.0\pm 0.1 \; , \; g_\omega=17.0\pm 0.3 \; , \;
g_\phi=12.9\pm 0.2 \; . \label{ggm} \end{eqnarray}

More complex is the situation with $g_{\omega\rho\pi}$ coupling constant.
Assuming vector meson dominance,  $\omega\rho\pi$ vertex appears in a number
of processes. For example, $V\to\pi^0\gamma$ decay proceeds via

\begin{center}
\begin{picture}(30000,15000)
\drawline\fermion[\S\REG](1000,7500)[300]
\drawline\fermion[\E\REG](\pbackx,\pbacky)[6000]
\drawline\fermion[\N\REG](1000,7500)[300]
\drawline\fermion[\E\REG](\pbackx,\pbacky)[6000]
\global\advance\pmidy by 1000
\put(\pmidx,\pmidy){$V$}
\drawline\fermion[\S\REG](\pbackx,\pbacky)[600]
\put(\pmidx,\pmidy){\circle*{600}}
\drawline\scalar[\SE\REG](\pmidx,\pmidy)[5]
\global\advance\pbackx by 1000
\put(\pbackx,\pbacky){$\pi^0$}
\global\advance\pbackx by 1000
\drawline\fermion[\NW\REG](\pfrontx,\pfronty)[300]
\drawline\fermion[\NE\REG](\pbackx,\pbacky)[4000]
\drawline\fermion[\SE\REG](\pbackx,\pbacky)[600]
\put(\pmidx,\pmidy){\circle*{600}}
\drawline\fermion[\SW\REG](\pbackx,\pbacky)[4000]
\global\advance\pmidx by -2000
\global\advance\pmidy by 700
\put(\pmidx,\pmidy){$V_1$}
\global\advance\pfronty by 400
\drawline\photon[\NE\REG](\pfrontx,\pfronty)[7]
\global\advance\photonbackx by 1000
\put(\photonbackx,\photonbacky){$\gamma$}
\end{picture}
\end{center}

\noindent and therefore $g_{V\pi\gamma}=\frac{g_{VV_1\pi}}{g_{V_1}}$.
So $g_{VV_1\pi}$ can be determined from $\Gamma(V\to\pi^0\gamma)$ and
$\Gamma(V_1\to e^+e^-)$. On the other hand, $\pi^0\to 2\gamma$ decay,
in the framework of the vector meson dominance, goes through

\begin{center}
\begin{picture}(30000,16000)
\drawline\scalar[\W\REG](10000,8000)[5]
\put(\pfrontx,\pfronty){\circle*{600}}
\global\advance\pmidy by 1000
\global\advance\pmidx by -1000
\put(\pmidx,\pmidy){$\pi^0$}
\drawline\fermion[\NW\REG](\pfrontx,\pfronty)[300]
\drawline\fermion[\NE\REG](\pbackx,\pbacky)[4000]
\drawline\fermion[\SE\REG](\pbackx,\pbacky)[600]
\put(\pmidx,\pmidy){\circle*{600}}
\drawline\fermion[\SW\REG](\pbackx,\pbacky)[4000]
\global\advance\pmidx by -1500
\global\advance\pmidy by 700
\put(\pmidx,\pmidy){$\rho$}
\drawline\photon[\NE\REG](\pfrontx,\pfronty)[7]
\global\advance\photonbackx by 1000
\put(\photonbackx,\photonbacky){$k_1$}
\drawline\fermion[\NE\REG](10000,8000)[300]
\drawline\fermion[\SE\REG](\pbackx,\pbacky)[4000]
\drawline\fermion[\SW\REG](\pbackx,\pbacky)[600]
\put(\pmidx,\pmidy){\circle*{600}}
\drawline\fermion[\NW\REG](\pbackx,\pbacky)[4000]
\global\advance\pmidx by -1500
\global\advance\pmidy by -700
\put(\pmidx,\pmidy){$\omega$}
\drawline\photon[\SE\REG](\pfrontx,\pfronty)[7]
\global\advance\photonbackx by 1000
\put(\photonbackx,\photonbacky){$k_2$}
\global\advance\photonbackx by 4000
\put(\photonbackx,8000){$+\; \; (k_1\longleftrightarrow k_2) \; ,$}
\end{picture}
\end{center}

\noindent and therefore
$$\Gamma(\pi^0\to 2\gamma)=\pi\alpha^2 m_\pi^3\left (\frac{g_{\omega
\rho\pi}}{g_\rho g_\omega} \right ) ^2 \; . $$
\noindent So $g_{\omega\rho\pi}$ can be extracted from $\Gamma(\pi^0 \to
2 \gamma)$, $\Gamma(\rho \to e^+e^-)$, and $\Gamma(\omega \to e^+e^-)$.

All these methods for $g_{\omega\rho\pi}$ determination give consistent
values:
\begin{eqnarray} &&
(11.7\pm 0.5){\rm GeV}^{-1} \; \; 
{\rm from} \; \Gamma(\rho\to e^+e^-)\; {\rm and} \; 
\Gamma(\omega\to\pi^0\gamma) \; ,\nonumber \\ &&
(12.6\pm 0.7){\rm GeV}^{-1} \; \; 
{\rm from} \; \Gamma(\omega\to e^+e^-) \; {\rm and} \; 
\Gamma(\rho^+\to\pi^+\gamma)  \; , \\ &&
(12.5\pm 0.9){\rm GeV}^{-1} \; \; 
{\rm from} \; \Gamma(\pi^0\to 2\gamma) , \;  \Gamma(\omega\to\pi^0\gamma),
\; \Gamma(\rho^+\to\pi^+\gamma) \; , \nonumber \\ &&
(11.8\pm 0.5){\rm GeV}^{-1} \; \; 
{\rm from} \; \Gamma(\pi^0\to 2\gamma) , \;  \Gamma(\omega\to e^+e^-),
\; \Gamma(\rho\to e^+e^-) \; .\nonumber \label{gwrp1} \end{eqnarray}
But if the value of $g_{\omega\rho\pi}$ is extracted from the experimental
$\omega \to
3 \pi$ decay width, assuming that this decay proceeds through $\omega \to
\rho \pi$ intermediate state and that the $g_{\rho\pi\pi}$ coupling constant
is determined from the $\Gamma(\rho \to 2\pi)$, one obtains 
\begin{eqnarray}
g_{\omega\rho\pi} = (14.3\pm 0.2)~{\rm GeV}^{-1} \; .
\label{gwrp} \end{eqnarray}
\noindent This is closer to the chiral model prediction \cite{Rajev}
$$g_{\omega\rho\pi}=\frac{3g_{\rho\pi\pi}^2}{8\pi^2f_\pi}
\approx 14.9~{\rm GeV}^{-1} \; ,
\; \; f_\pi \approx 93 {\rm MeV} \; . $$

$g_{\omega\rho\pi}$ can be estimated also from QCD sum rules \cite{QCD1,
QCD2} with the result $g_{\omega\rho\pi}=(16\pm2){\rm GeV}^
{-1}$ \cite{QCD2}.  

In Monte-Carlo simulation, we have used (\ref{gwrp}) and $g_{\phi\rho\pi}
\approx -0.81 {\rm GeV}^{-1}$, which follows from $\Gamma(\phi \to 3\pi)$.
These values assume only $\rho$-pole decay mechanism for $\omega \to 3\pi$
transition, without a possible $\omega \to 3\pi$ contact term 
\cite{conterm}. In fact, the uncertainty in these coupling constants
does not effect significantly the Monte-Carlo estimates for
the detection efficiencies.

\subsection{$\phi$--$\rho$--$\omega$ mixing contribution}
$\phi \to \omega \pi$ vertex is forbidden by $G$-parity, which is negative
for all three particles. Another way to see this is to notice that 
it is impossible to construct isospin invariant trilinear coupling between
$\phi$ and $\omega$ isosinglets and isovector pion.

But the isospin symmetry breaks due to electromagnetic effects and mass
difference between u and d quarks. As a result, pure isospin eigenstates
$\omega _I$ and $\rho _I$ mix and cease to be mass
eigenstates. Instead we will have a nondiagonal mass matrix
$$\left ( \begin{array}{cc} \rho_I & \omega_I  \end{array} \right )
\left ( \begin{array}{cc}  z^0_\rho & z_{\omega \rho} \\
z_{\omega \rho} & z^0_\omega \end{array} \right )
\left ( \begin{array}{c} \rho_I \\ \omega_I  \end{array} \right ) \; . $$
\noindent Mass eigenstates (physical $\rho$ and $\omega$ mesons) are
mixtures of isospin eigenstates, which for small 
$\rho$--$\omega$ mixing angle
$\epsilon_{\omega \rho}$ ($\; \sin{\epsilon_{\omega \rho}}\approx 
\epsilon_{\omega \rho}$) look like
$$ \begin{array}{c} \rho=\rho_I -\epsilon_{\omega \rho} \omega_I \\
\omega=\omega_I +\epsilon_{\omega \rho} \rho_I \end{array} \; . $$
\noindent The mixing angle $\epsilon_{\omega \rho}$ is determined from 
the condition
$$\left ( \begin{array}{cc} \rho_I & \omega_I  \end{array} \right )
\left ( \begin{array}{cc}  z^0_\rho & z_{\omega \rho} \\
z_{\omega \rho} & z^0_\omega \end{array} \right )
\left ( \begin{array}{c} \rho_I \\ \omega_I  \end{array} \right ) =
\left ( \begin{array}{cc} \rho & \omega  \end{array} \right )
\left ( \begin{array}{cc}  z_\rho & 0 \\
0 & z_\omega \end{array} \right )
\left ( \begin{array}{c} \rho \\ \omega  \end{array} \right ) \; , $$
\noindent which gives in the first order in $z_{\omega \rho}$ and 
$\epsilon_{\omega \rho}$ \cite{Connell} :
$$z_\rho \approx z^0_\rho \; , \; \; z_\omega \approx z^0_\omega \; , \; \;
\epsilon_{\omega \rho} = \frac{z_{\omega \rho}}{z_\omega - z_\rho} \; . $$
\noindent Here \cite{Connell} $z_V=(M_V-i\frac{\Gamma_V}{2})^2\approx M_V^2-
iM_V\Gamma_V$ is the resonance complex mass square. So
\begin{equation}
\epsilon_{\omega \rho} = \frac{z_{\omega \rho}}{M_\omega^2-M_\rho^2-
i(M_\omega\Gamma_\omega -M_\rho\Gamma_\rho)} \; .
\label{epsi} \end{equation}
\noindent As for $\rho$--$\omega$ mixing amplitude $z_{\omega \rho}$, we will
take the value
\begin{equation}
z_{\omega \rho}=(-3800 \pm 370) {\mathrm MeV}^2 \; ,
\label{zrw} \end{equation}
\noindent which follows from the pion form factor studies \cite{Connell}.

Since $\rho_I=\rho+\epsilon_{\omega \rho} \omega$, the presence of the 
$\phi \to \rho_I \pi$ vertex will induce an effective $\phi \to \omega \pi$ 
vertex with a coupling constant
$
g^{(\omega \rho)}_{\phi \omega \pi} \approx \epsilon_{\omega \rho} 
g_{\phi \rho \pi} \; .
$

The described above picture of $\rho$--$\omega$ mixing can be trivially 
generalized
to include mixings with the $\phi$ meson also \cite{Bramon}:
$$\begin{array}{c} \rho=\rho_I -\epsilon_{\omega \rho} \omega_I -
\epsilon_{\phi \rho} \phi_I \\
\omega=\omega_I +\epsilon_{\omega \rho} \rho_I -
\epsilon_{\omega \phi} \phi_I \\
\phi=\phi_I +\epsilon_{\phi \rho} \rho_I +
\epsilon_{\omega \phi} \omega_I
\end{array} \; . $$
\noindent Due to $\phi$--$\rho$ mixing another contribution to the
$\phi \to \omega \pi$ transition will arise
$
g^{(\phi \rho)}_{\phi \omega \pi} \approx \epsilon_{\phi \rho}
g_{\omega \rho \pi} \; .
$
Because $|g_{\omega \rho \pi}| \gg | g_{\phi \rho \pi}|$, this latter
contribution turns out to be of the same order as
$g^{(\omega \rho)}_{\phi \omega \pi}$. So for the
$g_{\phi \omega \pi}$ coupling constant we will use
\begin{equation}
g_{\phi \omega \pi}=g^{(\omega \rho)}_{\phi \omega \pi}+
g^{(\phi \rho)}_{\phi \omega \pi}=
\epsilon_{\omega \rho} g_{\phi \rho \pi}+
\epsilon_{\phi \rho} g_{\omega \rho \pi} \; .
\label{gfwp} \end{equation}

\noindent The corresponding contribution in $e^+e^- \to \pi^0 \pi^0 \gamma$
is then determined by equation (7).

The $\phi$--$\rho$ mixing parameter $\epsilon_{\phi \rho}$ can be extracted 
from the $e^+e^-\to \pi^+ \pi^-$ experimental data near the $\phi$-meson
\cite{Bramon}:
\begin{equation}
\epsilon_{\phi \rho}=Z\left ( \frac{M_\rho^2 g_\phi}{M_\phi^2 g_\rho}
\right )\frac{M_\phi \Gamma_\phi}{M_\phi^2-M_\rho^2+iM_\phi
\Gamma_\rho(M^2_\phi)} \; , 
\label{epsfr} \end{equation}
\noindent where $Z$ is experimentally measured interference magnitude
\cite{ND}. Taking average values $Re(Z)=(8\pm 2)\cdot 10^{-2}, \;
Im(Z)=-(3.5 \pm 1.3)\cdot 10^{-2}$ from \cite{ND}, we obtain
\begin{equation}
\epsilon_{\phi \rho}=(0.72 \pm 0.18 )\cdot 10^{-3}
-i(0.87 \pm 0.32)\cdot 10^{-3} \; .
\label{epsfrv} \end{equation}

\subsection{$\rho^{\, \prime}$ meson contribution}
No definite theoretical predictions exist for $\rho^{\, \prime}(1465)$ meson
contribution, nor it is definitely known whether the only 
one radial excitation
gives significant contribution in this energy region. We will use the 
following parameterization for this part of the $g(x)$ function
\begin{eqnarray} &&
 g^{(\rho^{\, \prime},\omega)}(x)=Re^{i\xi}k_\rho 
 \frac{s}{s-M_{\rho^{\, \prime}}^2+iM_{\rho^{\, \prime}}\Gamma_{\rho^{\,
\prime}}} \, \frac{1}{1-x+\gamma_{\omega}}
 \nonumber \\ &&
 \gamma_\omega=\frac{1}{s}(m^2-M_\omega^2+iM_\omega\Gamma_\omega),
 \label{gxrp} \end{eqnarray}
\noindent where $k_\rho=\pi\alpha M_\rho^2 \, \frac{g_{\rho \omega \pi}
g{\omega \pi \gamma}}{g_\rho}$ corresponds to the $\rho$-meson contribution.
$R$ and $\xi$ parameters were determined by fitting $\sigma(e^+e^- \to 
\omega \pi)=\frac{\sigma(ee \to\omega \pi\to\pi^0\pi^0\gamma)}
{Br(\omega\to\pi^0\gamma)}$ cross section to higher energy
experimental data from \cite{ND}. The fit gives the following values:
\begin{equation}
R=0.72 \pm 0.03 \; , \; \; \xi=3.25 \pm 0.08 \; .
\label{rksi} \end{equation}

Our results of this fit confirms the conclusions of \cite{Clegg} that the 
$\rho^{\, \prime}(1465)$ meson contribution is sufficient to describe the
existing experimental data in the $1.05 \div 1.6~{\rm GeV}$ energy range.

Note that we use the PDG \cite{PDG} values $M_{\rho^\prime}=1.465~{\rm GeV}$
and $\Gamma_{\rho^\prime}=0.31~{\rm GeV}$. To simulate roughly a threshold 
effect due to the dominant $\rho\pi\pi$ decay of the $\rho^\prime$ meson,
we have assumed, as in \cite{Buon}, that the $\rho^\prime$-meson width rises
linearly from $\sqrt{s}=0.8~{\rm GeV}$ up to $\sqrt{s}=M_{\rho^\prime}$,
remaining constant above.

\subsection{Theoretical prediction}
\begin{figure}[htb]
  \begin{center}
\mbox{\epsfig{figure=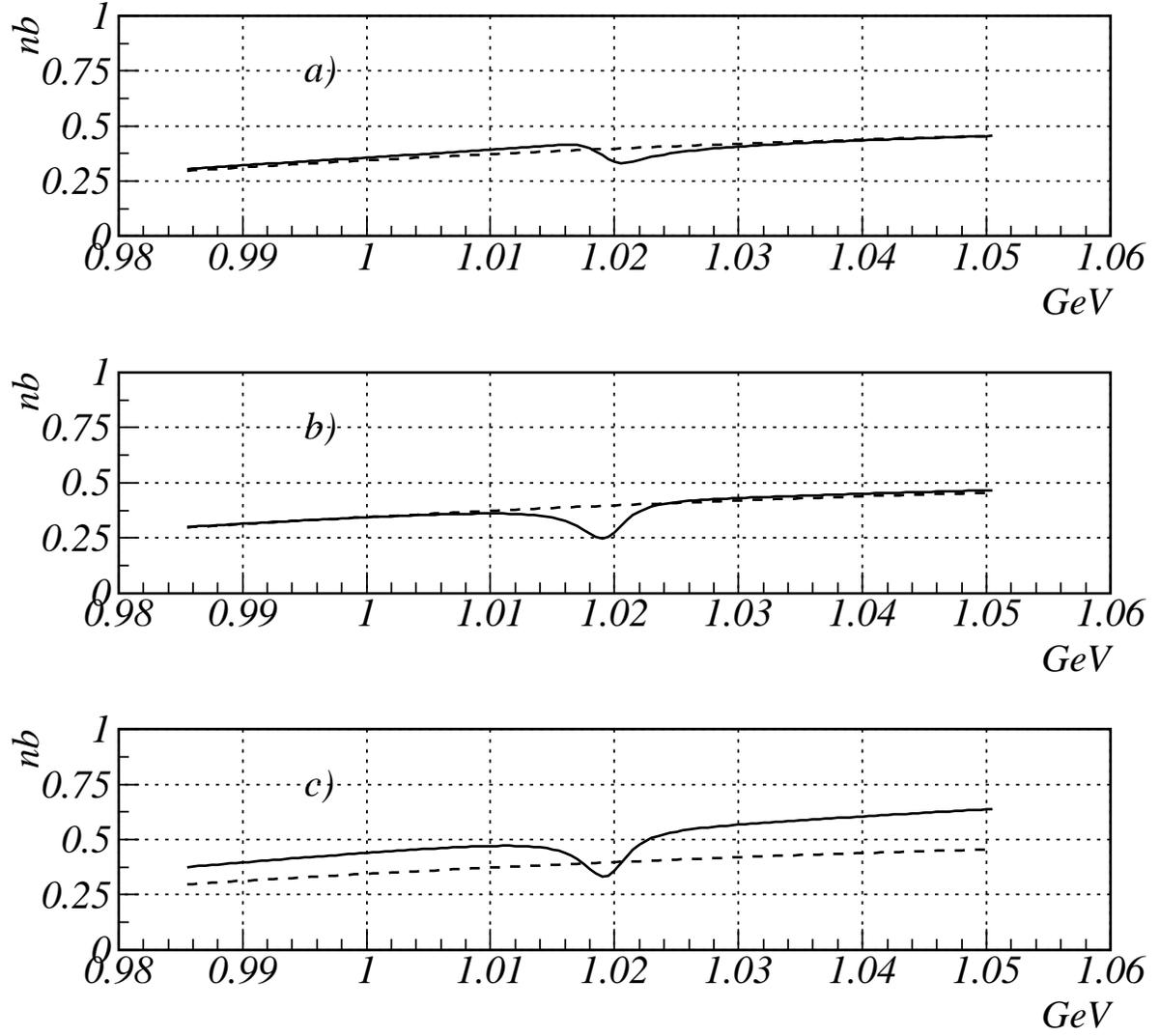
                               ,  height=14.5cm}}
   \end{center}
\caption {Theoretical predictions for $\sigma(ee \to \omega \pi^0
\to \pi^0 \pi^0\gamma)$. Dashed line - $\rho$-meson contribution only.
Solid line: a) $\phi$-meson contribution through $\phi \to \rho \pi^0 \to
\pi^0 \pi^0 \gamma$ is added. b) $\phi \to \omega \pi^0$ transition
due to $\rho$--$\omega$--$\phi$ mixing is also included in the $\phi$-meson
contribution. c) $\rho^\prime$-meson contribution is added to the
above ones.}
\label{Fig1}
\end{figure}

Using the estimates for various coupling constants given above,
the total cross section can be evaluated for $e^+e^-\to \pi^0\pi^0\gamma$
reaction. The result is shown on Fig.\ref{Fig1}.
Note the significant interference effect near $\phi$ meson, although the
$\phi$-meson mediated amplitude itself remains small and corresponds to
$Br(\phi\to\rho\pi\to\pi^0\pi^0\gamma) \approx 10^{-5}$, in consistence
with \cite{Ivanchenko}. The $\phi$--$\rho$--$\omega$ interference contribution
into $\phi \to \omega \pi$ transition corresponds to
Br($\phi\to\omega\pi) \approx 5.7 \cdot 10^{-5}$ in excellent agreement with
the recent experimental result \cite{PhiWpi} $(4.8^{+1.9}_{-1.7}\pm 0.8)
\cdot 10^{-5}$, although it should be beared
in mind that there may exist also other sources for this transition
\cite{Achasov}

\section{Detector and experiment}
\begin{figure}[htbp]
\centerline{\epsfig{figure=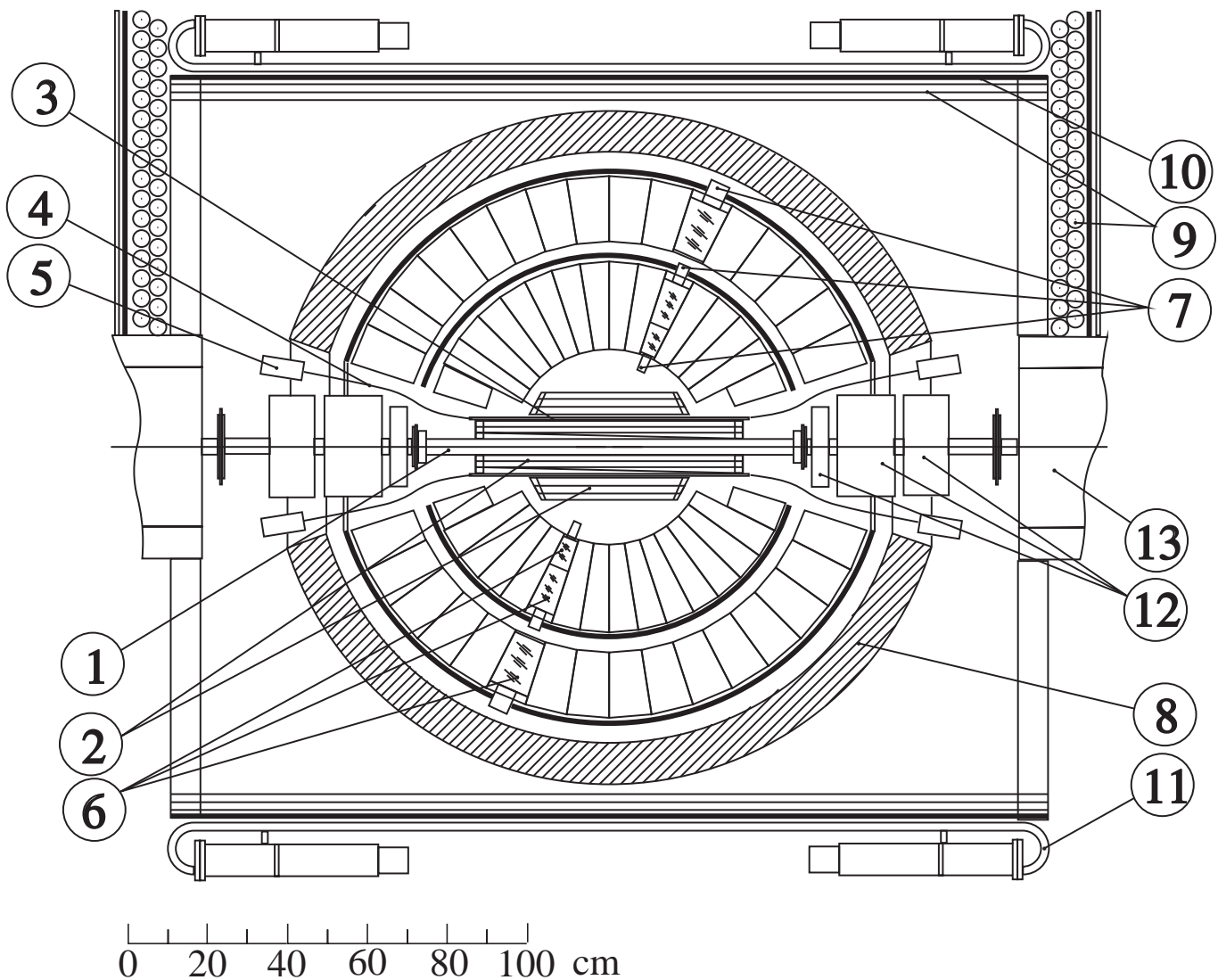
                               ,  height=14.5cm}}
\caption{ \label{SND}
  Detector SND --- view along 
   the beam; 1---beam pipe, 2---drift
   chambers, 3---scintillation counters, 4---fiber lightguides, 5---PMTs,
   6---NaI(Tl) counters, 7---vacuum phototriodes, 8---iron absorber,
   9---streamer tubes, 10---1~cm iron plates, 11---scintillation counters,
   12---magnetic lenses, 13---bending magnets
   }
   \end{figure}

SND is a general purpose nonmagnetic detector.  
This new detector combines advantages of its predecessor ND detector 
\cite{NDNIM} and famous Crystal Ball detector \cite{CrBall}, that is 
a good uniformity over the solid angle due to the spherical shape, 
a good $e/\pi$ and $\gamma/K_L$ separation due to multilayer 
structure of the electromagnetic calorimeter, and high hermeticity.
The main part of the SND detector (Fig.\ref{SND})
is a 3-layer, spherical,  highly
granulated, NaI(Tl) electromagnetic calorimeter \cite{SNDCAL}.
Tracking system, located in the detector center, consists of two 
coaxial cylindrical drift chambers. In the  radial
gap between them a 5~mm thick  plastic scintillation counter
with a wavelength shifter  fiber readout is placed.
From outside the calorimeter is covered by a thick (12~cm)
iron absorber, which attenuates
remnants of electromagnetic showers. An outer muon/veto system is located
outside the absorber. It consists of  sheets of plastic scintillator and 
streamer tubes.

The detailed description of the SND detector can be found elsewhere
\cite{SND,SNDPR,BEGEXP}. Here we repeat only the main points, relevant
to this study.

\subsection{Calorimeter}
Each layer of the calorimeter includes 520--560
crystals of eight different shapes. Most of the crystals have shapes of
truncated tetrahedral pyramids.  The total solid angle covered
by the calorimeter is equal to 0.9 of 4$\pi$.  The remaining space is
occupied by magnetic structure elements of the storage
ring, mainly by quadrupole lenses. Pairs of counters of the first two
layers with thickness of 3$X_0$ and 5$X_0$ respectively
are sealed together in common containers made of thin (0.1 mm) aluminum foil.
In order to improve light collection efficiency and to separate one
crystal from another, each crystal
is wrapped in aluminized mylar. The
gaps between adjacent crystals do not exceed 0.5~mm. The r.m.s. value
of the nonuniformity of the light collection efficiency along the
crystals is less than 3~\%. All the containers are fastened to 5 mm
aluminum supporting hemispheres. The outer layer of 6$X_0$ thick counters
has a similar design. Thus the total thickness of the calorimeter
is 13.5$X_0$ (35 cm) of NaI(Tl).

As photosensitive devices for the calorimeter counters vacuum
phototriodes are used \cite{Beschastnov}. The quantum efficiency of their
photocathodes is about 15~\%, average gain is  10, and  light collection
efficiency for individual counters is  about 10~\%. Signals from
phototriodes are amplified by charge sensitive preamplifiers located
directly on the counters. Output signals
are carried to shaping amplifiers via  20 m long twisted cables.
For the trigger needs the calorimeter crystals are logically organized
into ``towers''. A tower consists of counters
located within the same $18^\circ$ interval in 
polar and azimuthal directions
in all three layers. The number of counters in a tower is 12 at large
angles and 6 in the regions close to the beam. In addition to signals
from individual counters each tower produces an analog total energy 
deposition signal and  two trigger signals. In order to equalize 
contributions from
different counters into the total energy deposition signal and to obtain
equal energy thresholds for trigger signals
over the whole calorimeter, all shaping amplifiers
are equipped with computer controlled attenuators, allowing to adjust
channels gain in steps of 1/255.
The resulting electronics noise is close to 0.3~MeV (r.m.s.).
The dependence of the calorimeter energy resolution on photon energies 
was fitted as \cite{Bhabha}
\begin{equation}
   \sigma_E/E(\%) = {4.2\% \over \sqrt[4~]{E(\mathrm{GeV})}} \; .
\nonumber \end{equation}

The calorimeter energy resolution is determined mostly by the amount of
passive material inside the calorimeter.  Noticeable contribution  comes 
also from the shower energy leakage due to  limited thickness of NaI(Tl) and
gaps between crystals. It turned out that the nonuniformity 
of the light
collection efficiency in the NaI(Tl) crystals of the inner calorimeter
layer contributes significantly into energy resolution. The simulation
of calorimeter response agrees reasonably well with experiment after
these effects are included into simulation program.

Absolute energy calibration of the calorimeter is performed by using 
cosmic muons \cite{AchasovM} and Bhabha scattering \cite{Bhabha}.
The latter process, together with the two-photon annihilation reaction,
was also used for the luminosity measurements. The systematic error
in the integrated luminosity determination is about 3\%.

After the calorimeter calibration with $e^+e^- \rightarrow e^+e^-$ events,
the photon energies turned out to be biased by about $1\%$.
In order to compensate for this bias, the calibration coefficients for 
photons were corrected accordingly.

\subsection{Tracking system}
The tracking system consists of two cylindrical drift chambers. The length
of the chamber closest to the beam is 40~cm, its inner and outer
diameters are 4 and 12~cm respectively. The corresponding dimensions of the
outer chamber are 25, 14 and  24~cm. Both chambers are divided
into 20 jet-type cells in azimuthal plane. Each cell contains 5 sense
wires. The longitudinal coordinate is measured by
the charge division method with an accuracy of 3~mm. In addition,
a cathode strip readout for inner and outer layers provides the
improvement of the latter value to 0.5~mm. The outer drift chamber improves
pattern recognition for multiparticle events. The overall
angular resolution of the drift chamber system is 0.5 and 1.7 degrees in
azimuthal and polar directions respectively. The impact parameter
resolution is 0.5~mm. The solid angle coverage for the inner chamber 
is 96~\% of 4$\pi$.

\subsection{Muon detector}
The SND muon system consists of plastic scintillation counters and strea\-mer 
tubes \cite{Streamer}. It is intended mainly for cosmic background
suppression for the events without charged particles. The probability of its
triggering by the events of the reaction $e^+ e^- \to \gamma\gamma$
at the maximum available energy is less than 1~\%.
Muons of the process $e^+e^- \to \mu^+\mu^-$, starting from
the beam energy about 500~MeV, penetrate through the absorber and
hit the muon detector. 

\subsection{Trigger}
The three-level trigger of the detector  selects  events
of different types: events with photons only, events with
charged particles, and cosmic muon events for the detector calibration.
The drift chamber first level trigger (FLT) searches for tracks in the drift
chambers with impact parameter $\Delta r < 20$~mm. The calorimeter FLT
uses the total energy deposition in the calorimeter and double coincidences
of calorimeter towers with a threshold of 30~MeV.

Detector electronics provides also signals for a second level trigger.
But at present the second level trigger proved to be unnecessary and
was not implemented, although a slot is left for it in the electronics.

The third level trigger (TLT) is implemented as a special fast computer code
on a main data acquisition computer. It checks events before recording them
on tape, rejects cosmic events with charged trigger, and
suppresses beam-related background using $z$-coordinates of tracks,
measured by drift chambers. In addition TLT identifies collinear events of
Bhabha scattering and 2$\gamma$ annihilation. These events are used for
monitoring the collider luminosity.

\subsection{Data acquisition system}
SND data acquisition system is based on the fast electronics modules
KLUKVA \cite{KLUKVA}, developed in BINP specially for purpose of the
detectors CMD-2, KEDR and SND.

Analog signals from the detector come to the front-end amplifiers and
shapers, located near the detector. Then the signals are transmitted
via screened twisted pair cables to the digitizing modules (ADC, TDC, etc) 
in KLUKVA crates.  Logical signals from  discriminators in KLUKVA crates
are collected by the first level trigger interface modules. After 1~$\mu s$,
needed for FLT decision, FLT generates a signal to start digitizing.
The contents of KLUKVA digitizing modules are read into the RAMs of Output 
Processor modules (OP), located in KLUKVA crates. This procedure finishes 
in 120~$\mu s$. There are
two independent RAMs for events in each OP and one event can be stored in
digitizing modules, such a fast buffering greatly decreases dead time.
The total dead time in KLUKVA is 200~$\mu s$ per event selected by FLT.
The data from OP RAMs are read by the VAX server 3300 through the CAMAC
interface modules with the rate of 2 ms/event. The TLT program can process
events with a maximum rate of 45Hz.
Finally the processed events are put on 8~mm 5~GB EXABYTE tapes.

\subsection{Experiment PHI96}
A description of the PHI96 experiment was published in \cite{EXP96}.
The PHI96 experiment was carried out from February 1996 
up to  January 1997. Seven successive scans 
were performed in 14 energy points in the range $2E$ from 980 to 1044~MeV.

The data sample, which was analyzed, corresponds to an integrated luminosity
of $4.5\, pb^{-1}$, collected by SND in the narrow energy interval near 
the $\phi$-meson. Estimated number of produced $\phi$-mesons equals to
$8.3 \cdot 10^6$.

\section{Event selection}
While studying the channel
\begin{equation}
e^+e^-\to\omega\pi^0\to\pi^0\pi^0\gamma
\label{eewpi} \end{equation}
\noindent one should be aware about the possible background
from the following processes
\begin{equation}
e^+e^-\to\phi\to\eta\gamma\to 3 \pi^0\gamma \; ,
\label{phietg} \end{equation}
\begin{equation}
e^+e^-\to\phi\to K_S K_L\to {\rm neutral \; \; \; particles} \; .
\label{phikk} \end{equation}
A primary selection of the $e^+e^-\to\omega\pi^0\to\pi^0\pi^0\gamma$
candidates was done according to following criteria:
\begin{itemize}
\item the event must contain exactly 5 photons in the calorimeter and have 
no charged particles,
\item the azimuthal angle of any final photon lies within the interval 
$27^o<\theta<153^o$,
\item the total (normalized over $2E$) energy deposition of final photons is
in the range $0.8 \le E_{tot}/2E \le 1.1$,
\item the normalized full momentum of the event ($P_{tot}/2E$) is less 
than 0.15.
\end{itemize}
The two latter conditions eliminate the main part of the  background, 
originated 
from the $\phi \to K_LK_S$ decay, but they 
do not help much against another
background, coming from the process (\ref{phietg}).

\begin{figure}[htb]
\centerline{\epsfig{figure=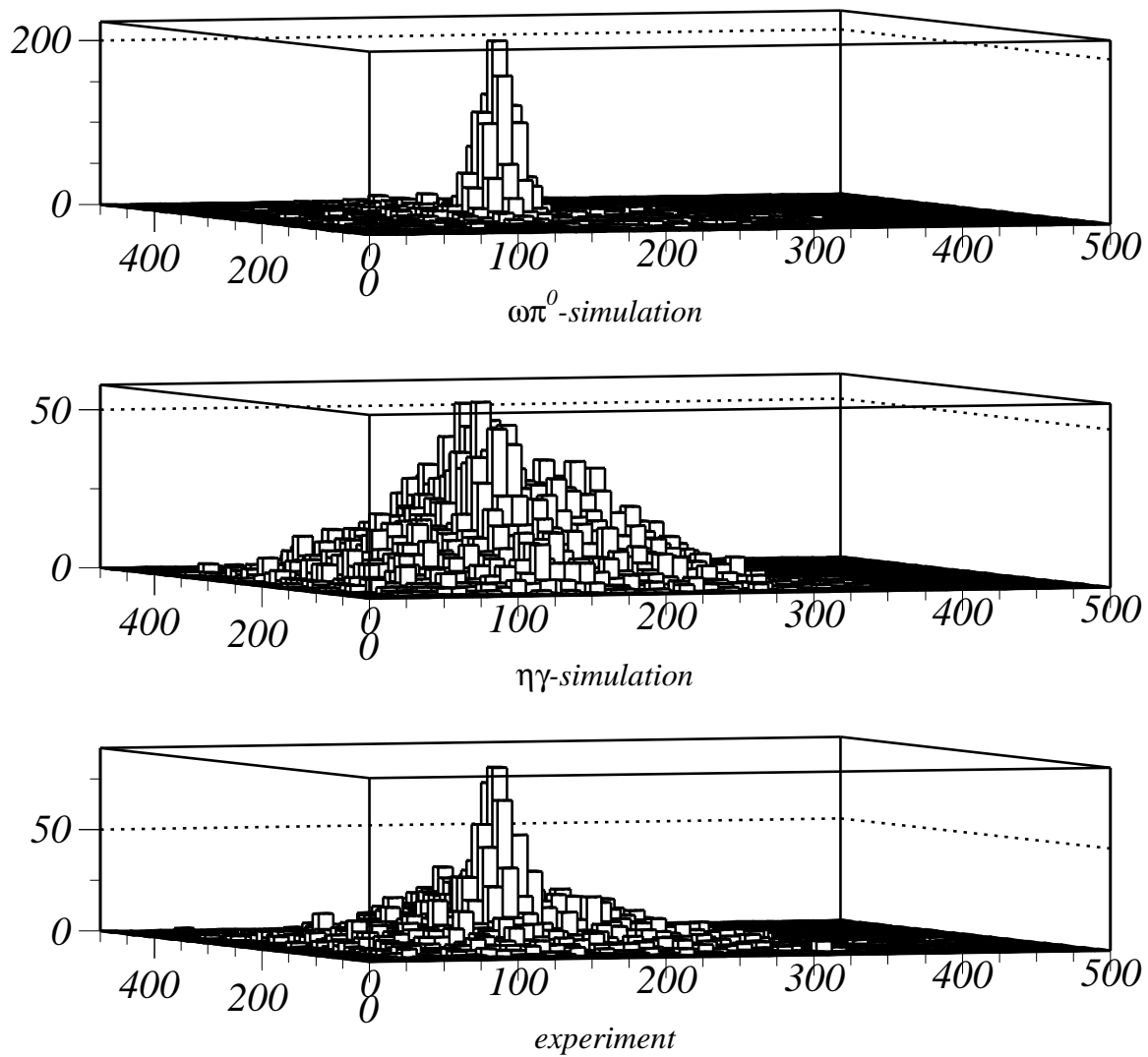
                               ,  height=14.5cm}}
\caption {Distributions $M_{24}$ versus $M_{35}$ plus 
$M_{25}$ versus $M_{34}$ for MC simulation and experiment.}
\label{Fig3}
\end{figure}

It is interesting to check whether we really have two $\pi^0$-s in our
5-photon events. It is unlikely for the most energetic photon to come
from $\pi^0$ decay in either (\ref{eewpi}), (\ref{phietg})
or $e^+e^-\to K_LK_S\to 2\pi^0 K_L$
reactions. It is also not likely for $\pi^0$ to produce the two softest
photons.  If we discard the corresponding combinations, only two possibilities
(2,4),(3,5) and (2,5),(3,4) are left for the photons to compose 
$\pi^{0,}$s (photons are arranged according to their energy, 
the most energetic
being the first one). A combined 2-dimensional plot of invariant masses 
of these photon pairs ($M_{24}$ versus $M_{35}$ plus
$M_{25}$ versus $M_{34}$) is shown in Fig.\ref{Fig3}. 
We clearly see that our 5-photon events are
predominantly $2\pi^0$ events also. Besides, Fig.\ref{Fig3} illustrates that 
background (\ref{phietg}) produces a wider distribution.
This can be used for the rejection of this background. For this purpose,
a kinematic fit was performed for each 5-photon event under the
assumption that there are two $\pi^{0,}$s in the final state and 
energy-momentum balance holds within the experimental accuracy. 
A $\chi^2$ of
this fit ($\chi^2_{\pi^0\pi^0\gamma}$) can be used for the background 
rejection.

\begin{figure}[htb]
\centerline{\epsfig{figure=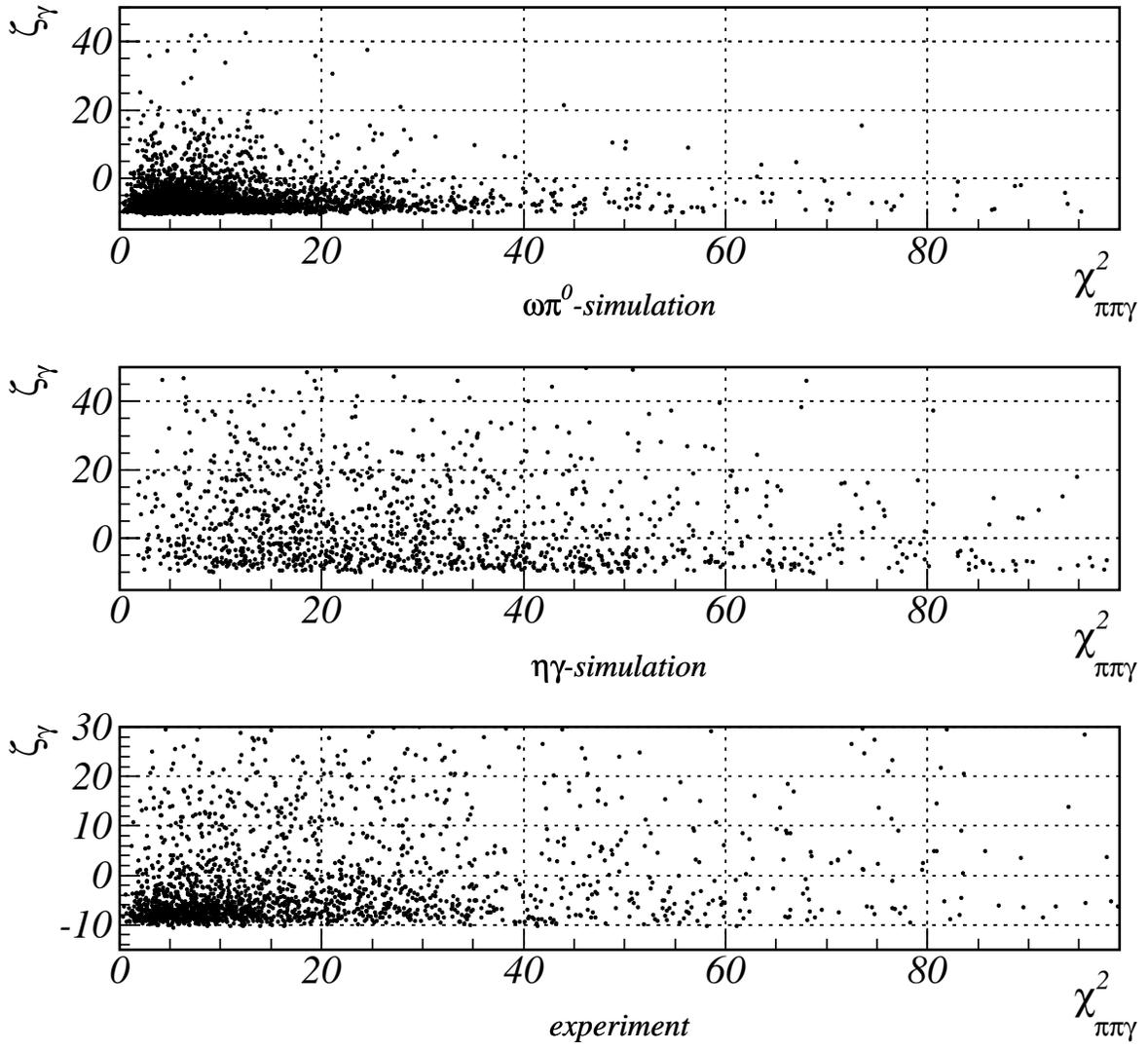
                               ,  height=14.5cm}}
\caption {$\zeta_\gamma$ versus $\chi^2_{\pi^0\pi^0\gamma}$ distributions
for MC simulation and experiment.} 
\label{Fig4}
\end{figure}

Background from (\ref{phietg}) simulates $\pi^0\pi^0\gamma$
events mainly due to loss of photons through the openings in
the calorimeter around detector poles and/or merging of close photon pairs.
When photons merge in the calorimeter, the corresponding electromagnetic
shower is, as a rule, broader in transverse direction than the  
electromagnetic showers from the single photons. This circumstance
can be used to discriminate against merged photons and so against 
a great deal of 
background (\ref{phietg}). The corresponding parameter ($\zeta_\gamma$)
is described in \cite{TrEnProf}. A 2-dimensional distributions of our events
in the $\chi^2_{\pi^0\pi^0\gamma},\zeta_\gamma$ plane, as well as 
Monte-Carlo simulated (\ref{eewpi}) signal events and (\ref{phietg}) 
background events (Fig.\ref{Fig4}), indicate that our 
signal events are almost completely
bound in the $\zeta_\gamma<20, \; \chi^2_{\pi^0\pi^0\gamma}<40$ area.
This was confirmed by 
experimental events outside of the $\phi$-meson, where there should 
be no background from the process (\ref{phietg}).  

On the basis of these considerations, we have chosen the following 
two sets (Cut I and Cut II) of selection criteria for the 
channel (\ref{eewpi})  separation
(in addition to the primary selection rules, described above):
\begin{itemize}
\item the normalized full momentum of the event is less than 0.1 (for Cut I),
\item there are two $\pi^0$-mesons in the event, that is, one can find two
distinct pairs of photons with invariant masses within $\pm30$ MeV from the 
$\pi^0$ mass,
\item $\chi^2_{\pi^0\pi^0\gamma}$, the $\chi^2$ of the kinematic fit, is less 
than 20 for Cut I, or is less than 40 for looser Cut II,
\item $\zeta_\gamma$, the parameter describing the transverse profile of the 
electromagnetic shower, is less than 0 for Cut I, or is less than 20 for 
Cut II.
\end{itemize}

One more source of background is a recently observed process 
\cite{EXP97,f0gamma}:
\begin{equation}
e^+e^- \to \phi \to f_0 \gamma \to \pi^0 \pi^0 \gamma \; .
\label{phif0g} \end{equation}
\noindent  The recoil mass
of the photon from this process is peaked at the $f_0$-meson 
mass and this peculiarity can be used to separate a great deal of
such events from the events of process (\ref{eewpi}).
We have chosen the $M_\gamma<700 {\mathrm MeV}$ condition as one more cut 
to select 
events from the process (\ref{eewpi}), where $M_\gamma$ stands for 
the photon recoil mass.

\begin{figure}[htb]
  \begin{center}
\mbox{\epsfig{figure=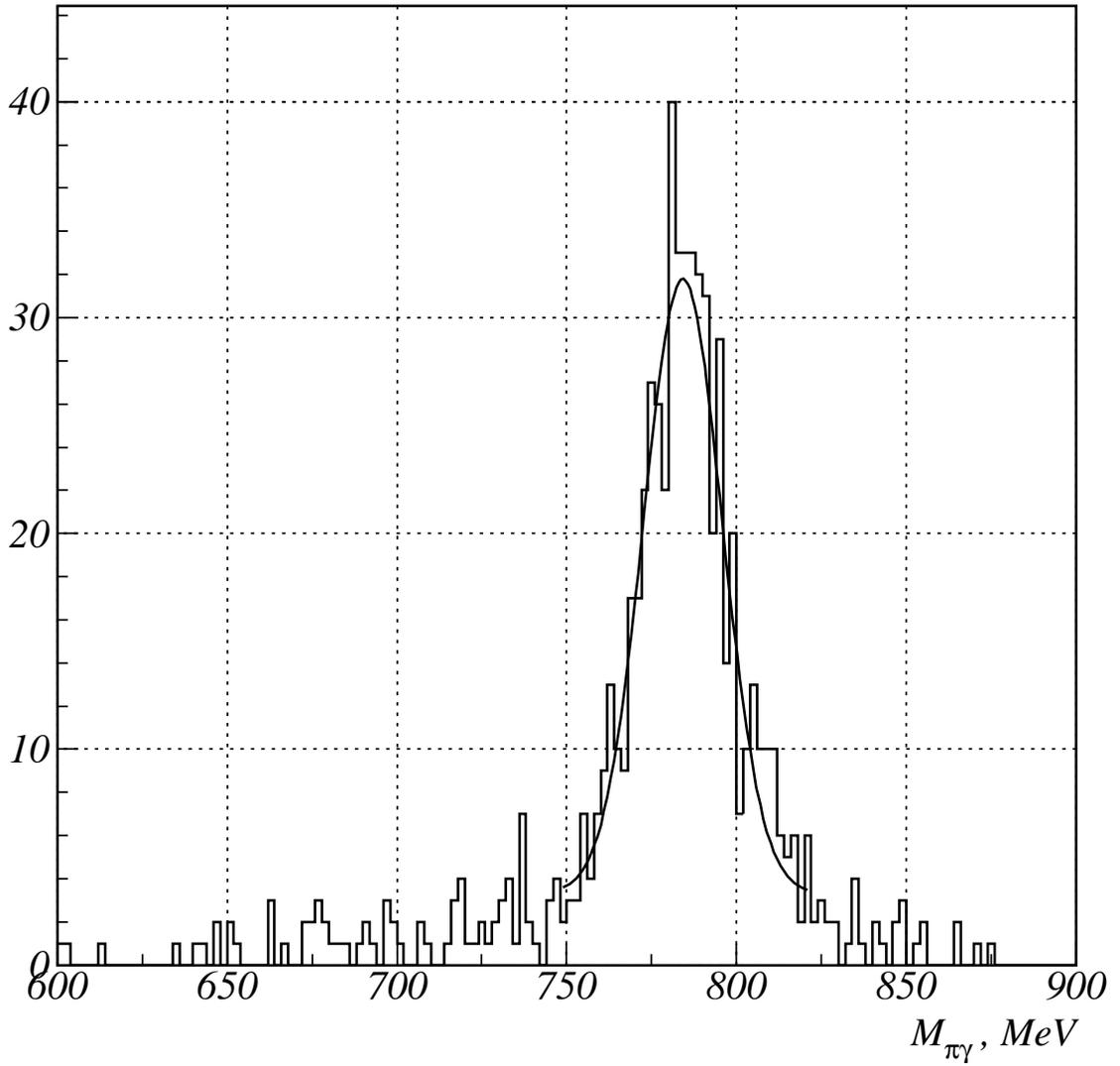
                               ,  height=14.5cm}}
   \end{center}
\caption {$M_{\pi^0\gamma}$ (invariant mass of $\pi^0\gamma$ nearest to 
$\omega$) distribution for Cut I.}
\label{Fig5}
\end{figure}

After applying these cuts, the $\omega$-meson peak is clearly seen in the 
invariant mass of $\pi^0$ and $\gamma$ (for each $\pi^0\pi^0\gamma$ event,
from two possible ($\pi^0,\gamma$) combinations, the one is taken, which has
$M_{\pi^0\gamma}$ closest to $M_\omega$), as it is illustrated by 
Fig.\ref{Fig5} (Cut I). Finally, to extract
channel (\ref{eewpi}), the $750{\rm MeV}<M_{\pi^0\gamma}<820{\rm MeV}$ 
condition was added to the above mentioned cuts. 

The $M_{\pi^0\gamma}$ histograms were 
fitted by  a function 
$$p_1\exp{(-\frac{(x-p_2)^2}{2p_3^2})}+p_4 \; .$$
\noindent The fit indicates a good agreement between the
experiment and simulation. For the 2356 ~MC events , which passed
Cut I, the fitted value of the constant term 
is $p_4=9.2 \pm 1.3$. Since in the 
$750{\rm MeV}<M_{\pi^0\gamma}<820{\rm MeV}$
interval we have 35 histogram bins, such background constant corresponds
to $322 \pm 46$  combinatorial background events,
that is about $14 \%$. The same conclusion follows also for Cut II.

For the 560  experimental events which passed Cut I, the following background
constant was obtained: $p_4=3.3 \pm 0.9$,
which correspond to $116 \pm 32$  background
events. Subtracting the expected combinatorial background of
$76 \pm 11$, we can estimate the residual
background from (\ref{phietg}), (\ref{phikk}), (\ref{phif0g}),
as $40\pm 34$ events ($\sim 7 \%$).
For Cut II this background was found to be twice higher, that is
$133\pm 38$ background events from the selected
864 experimental events ($\sim 15 \%$).

On the other hand, Monte Carlo studies indicate the following probabilities
for events from the main background sources to pass 
Cut I: $(2.7 \pm 0.3)\cdot 10^{-4}$ for 
process (\ref{phietg}), and $(2.9 \pm 0.2)\cdot
10^{-2}$ for 
process (\ref{phif0g}). This corresponds to 
$28 \pm 3$ background events from  
process (\ref{phietg}) and $22 \pm 4$  from  
process (\ref{phif0g}).  So the total number of background
events from these sources is estimated to be $50\pm 5$.
For Cut II we expect $116 \pm 8$ and
$27 \pm 5$ background events, respectively, or in total
$143 \pm 10$ events. These numbers are in a good agreement with those 
obtained from the above mentioned fitting of the $\omega$-meson peak.

To check the accuracy of Monte Carlo simulation  the following procedure was
applied. For the above described Cut I and Cut II, one of the parameters
is released and the resulting experimental distribution $H_{EXP}$ for this
parameter is compared with the sum of MC predictions for the signal $H_{
\omega \pi^0}$ and expected backgrounds from (\ref{phietg}) and
(\ref{phif0g}) sources: $H_{MC}=k_{\omega \pi^0}H_{\omega \pi^0}+
k_{\eta \gamma}H_{\eta \gamma}+k_{f^0\gamma}H_{f^0\gamma}$. The normalization
coefficients $k_{\eta \gamma}$ and $k_{f^0\gamma}$ are determined by the
total statistics of generated MC samples for this reactions. The normalization 
coefficient for the signal $k_{\omega \pi^0}$ is then determined from 
the condition that $H_{EXP}$ and $H_{MC}$
histograms have the equal total numbers of events. Note that 
any variation in the $k_{\omega\pi^0}$ coefficient indicates some
systematics and/or other background sources, not accounted for in the
comparison. For Cut I the averaged deviation turned out to be 3\%
and for Cut II -- 4\%.

\begin{figure}[htb]
  \begin{center}
\mbox{\epsfig{figure=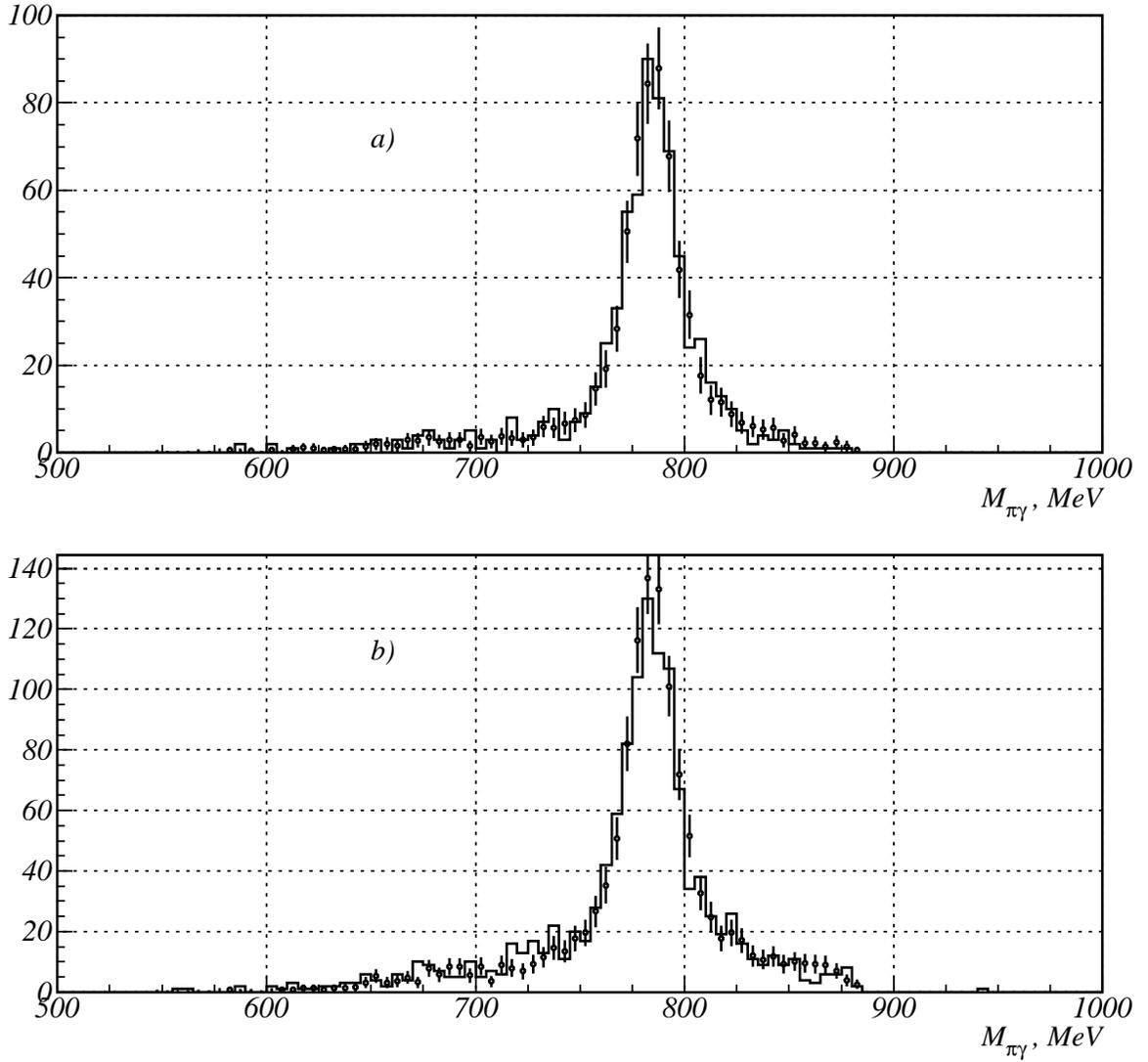
                               ,  height=14.5cm}}
   \end{center}
\caption {$M_{\pi^0\gamma}$ distributions : MC simulation (points with 
error bars) and experiment (histogram). 
a) for Cut I, and b) for Cut II.}
\label{Fig6}
\end{figure}

The distributions for all parameters used in event selections
show good agreement between $MC$ and experiment.
This indirectly indicates that the possible background from 
process (\ref{phikk}) is rather small and does not exceed $\sim 5 \%$.
As an example, on Fig.\ref{Fig6} we present $M_{\pi^0\gamma}$ distributions.

\section{Data analysis and results}
We assume the following parameterization for the visible (detection) cross
section $\sigma_v$:
\begin{eqnarray}
\sigma_v=\epsilon [1+\delta(s)]\sigma(s)+kb\sigma_B(s) \; ,
\label{csvis} \end{eqnarray}
\noindent where $\epsilon$ is the detection efficiency for the 
process (\ref{eewpi}), $\delta(s)$ accounts
for the radiative corrections, which are calculated according to the
standard procedure \cite{RC}, $\sigma_B(s)$ is a background cross section,
which was assumed to coincide with $\sigma(e^+e^-\to\eta\gamma\to
3\pi^0\gamma )$, $k$ is the background suppression factor, $b$ is described
below. Finally,
$\sigma(s)$ is a cross section of the process under investigation.
Because we are interested in $\sigma(s)$ in a narrow energy interval,
we have taken
\begin{eqnarray}
\sigma(s)=\left [\sigma_0+\sigma^\prime(2E-M_\phi)\right ]|R|^2, \; \;
R=1-Z\frac{m_\phi \Gamma_\phi}{s-M_\phi^2+iM_\phi\Gamma_\phi}.
\label{sgfit} \end{eqnarray}
The detection efficiency $\epsilon$ was calculated using
Monte-Carlo simulation in conditions of individual scans for various 
energies. Detection efficiencies do not show
any significant energy dependence.
So we have taken an efficiency value averaged over scans and energies, 
$\epsilon=(29.7\pm 0.25) \% $ (only statistical error is indicated) 
as a fair estimate for Cut II.

For a tighter Cut I some systematic errors could be expected. To estimate 
this systematics, we compared the numbers of rejected events for each 
parameter of Cut I in the above described $H_{EXP}$ and $H_{MC}$ 
distributions. The following correction factors 
($\epsilon_{MC}/\epsilon_{EXP}$) were obtained: 
$1.032 \pm 0.018$ for $P_{tot}/2E$, $1.015 \pm 0.023$ for 
$\chi^2_{\pi^0\pi^0\gamma}$, $1.074 \pm 0.028$
for $\zeta_\gamma$, $1.015 \pm 0.023$ for $M_\gamma$ and $0.98 \pm 0.028$ 
for $M_{\pi^0\gamma}$.
In total, the Monte Carlo simulation overestimates the detection efficiency
$1.12 \pm 0.06$ times,
if we assume that there are no correlations between the
used selection parameters. With this correction factor taken into account,
an average detection efficiency  $\epsilon=(20.7\pm 1.1) \% $ was obtained
for Cut I.

The background suppression factor $k$ was also assumed to be energy
independent. It was calculated using $\sim 2.15 \cdot 10^5$ simulated
events from the process (\ref{phietg}) and equals 
$(2.7\pm 0.3)\cdot 10^{-4}$ for Cut I
and $(1.10\pm 0.06)\cdot 10^{-3}$ for Cut II.

As we have seen above, for Cut I a relatively large fraction 
of background is
expected from process (\ref{phif0g}). Some
small amount of $\phi\to K_SK_L$ decay related background (\ref{phikk}) 
is also not excluded. To take into account these and other $\phi$-meson 
related backgrounds, the factor $b$ is introduced in (\ref{csvis}). It is 
assumed that the different energy dependencies of various background 
cross-sections is not
relevant at the present level of statistical accuracy and so they all can
be approximated by the $\sigma(e^+e^-\to\eta\gamma)$ behavior.

The fit results 
are given in the Table~\ref{tb1}.
\begin{table}[htb]
\begin{center}
\begin{tabular}{|c|c|c|}
\hline
parameter &  Cut I   &  Cut II \\
\hline
$\sigma_0$ (nb)& 0.61 $\pm$ 0.06 & 0.63 $\pm$ 0.05   \\
$\sigma^\prime$ (nb/MeV) & $(0.43\pm 0.24)\cdot 10^{-2}$  &
$(0.49\pm 0.21)\cdot 10^{-2}$ \\
Re($Z$) & $0.1\pm 0.1$ & $0.12\pm 0.08$ \\
Im($Z$) & $-0.19 \pm 0.15$ & $0.05 \pm 0.08$ \\
$b$ & $2.8\pm 1.9$ & $0.1\pm 0.3$  \\
\hline
$\chi^2/d.f.$ & 11.0/10& 14.5/10 \\
\hline
\end{tabular}
\end{center}
\caption
{Fitted parameters for the fit with the interference.}
\label{tb1}
\end{table}

Because of specific behavior of the interference effect, it can happen that
the resonant background cancels the interference dip and so mimics the
no-interference no-background situation, making it impossible to distinguish
between these two options during the fit. This is confirmed by the following
observation: If we assume no interference effects in the $\sigma(s)$ and just
take $\sigma(s)=\sigma_0+\sigma^\prime(2E-M_\phi)$, this linear
fit will result in the Table~\ref{tb2}. 
\begin{table}[htb]
\begin{center}
\begin{tabular}{|c|c|c|}
\hline
parameter &  Cut I   &  Cut II \\
\hline
$\sigma_0$ (nb)& 0.58 $\pm$ 0.05 & 0.64 $\pm$ 0.03   \\
$\sigma^\prime$ (nb/MeV) & $(0.44\pm 0.23)\cdot 10^{-2}$  &
$(0.58\pm 0.20)\cdot 10^{-2}$ \\
$b$ & $0.2\pm1.1$ & $0.2\pm 0.2$  \\
\hline
$\chi^2/d.f.$ & 12.1/12& 16.7/12 \\
\hline
\end{tabular}
\end{center}
\caption
{Fitted parameters for the linear fit. }
\label{tb2}
\end{table}

In spite of our considerations above, from which one could expect 
$b\approx 1$ for Cut II (and $b\approx 2$ for Cut I -- the expected 
background is indeed dominated by the process (\ref{phietg}) for Cut II, 
while for Cut I half of the background originates due to process 
(\ref{phif0g})), this fit indicates no-background for Cut II, so in fact
providing an indirect evidence in favour of the interference effect! 

As we see, the results for Cut I and Cut II are consistent 
after the detection efficiency for Cut I is corrected against estimated
systematics. Of course, we do not know if there is some part of
systematic errors left uncorrected for Cut I. On the other hand, for Cut II
more background is expected and we can neither very precisely estimate this
background (for example, the part coming from (\ref{phikk})) nor
subtract it during fit (note that all of the above fits
gave very large errors for the background constant $b$). 

Therefore we select as a fair estimates for the $\sigma_0$ and $\sigma^\prime$
parameters the mean values between Cut I and Cut II, and as systematic
errors we take the difference between them. Another part ($\sim 3\%$) of the
systematic error can arise from the luminosity measurement.

One more source of systematics is a possibility that a sixth false photon is
piled up on the five photon event due to the beam related background. 
This effect
is hard to simulate, so the MC efficiency is expected to be somewhat 
overestimated. To find the correction factor, the 6-photon events were 
investigated as described in the next section. This factor turned out to be
$1.04\pm 0.02$.
Our final results in this analysis, which include this correction factor, as
well as the estimated total systematic error, are

\begin{eqnarray} &&
\sigma_0 =  ( 0.64 \pm 0.08 ) \; {\mathrm nb} \nonumber \\ &&
\sigma^\prime = (0.48 \pm 0.27 )\cdot 10^{-2} \; {\mathrm nb/MeV} 
\label{results} \end{eqnarray}
\noindent As for
$Re(Z)$ and $Im(z)$, the still high level of background precludes 
an estimation of
these parameters. For Cut I, which is more pure against background,
$Im(Z)= -0.19 \pm 0.15$ is only one sigma effect.

We have also tried $\sigma(s)=\zeta\sigma_{th}(s)$ in a fit, where
$\sigma_{th}(s)$ is the theoretical prediction (including 
$\rho^\prime$-meson) for the reaction (\ref{eewpi}), 
discussed at the beginning. 
The fitted parameters are given in the Table 3. 
\begin{table}[htb]
\begin{center}
\begin{tabular}{|c|c|c|}
\hline
parameter &  Cut I   &  Cut II \\
\hline
$\zeta$ & $1.19 \pm 0.09 \; \; (1.54 \pm 0.12)$ & $1.29 \pm 0.08   
\; \; (1.68 \pm 0.11)$ \\
$b$ & $2.7\pm 1.0 \; \; (2.8 \pm 1.0$) & $1.1\pm 0.3 \; \; 
(1.2 \pm 0.3$) \\
\hline
$\chi^2/d.f.$ & 11.7/13 $\; \;$ (11.7/13) & 17.5/13 $\; \;$ (17.4/13) \\
\hline
\end{tabular}
\end{center}
\caption
{Fitted parameters for the fit with the theoretical cross-section.
The parameters for the fit without $\rho^\prime$ meson are also given
in the parenthesis. The indicated $\zeta$-values should be multiplied by the
correction factor $1.04\pm 0.02$, as discussed in the text. }
\label{tb3}
\end{table}

The obtained cross section is close enough to the theoretical prediction,
when $\rho^\prime$-meson contribution is also included, although 
the experimental cross section is somewhat ($\sim 25 \%$) higher. This
can indicate that more then one radial excitation of the $\rho$-meson
contributes or that the used higher energy $ee\to\omega\pi$ cross section
is subject to some systematics, which we do not take into account while
extracting the $\rho^\prime$ related parameters $R$ and $\xi$. Finally,
our parameterization for the $\rho^\prime$ contribution can also be
not quite adequate (for example, we had only roughly modeled  the energy
dependence of the $\rho^\prime$ width). In any case, the experimental
cross section can not be explained by only the $\rho$-meson tail, which
gives about 1.6 times lower result.

\section{Observation of the interference effect}

During the study of the reaction
$e^+e^-\to\omega\pi^0\to\pi^+\pi^-\pi^0\pi^0$ ,
the $\phi\to\omega \pi^0$ decay was observed for the first time with
the branching ratio  about $5\cdot 10^{-5}$ \cite{PhiWpi}.
The decay reveals itself as an interference wave on the nonresonant
cross section of the process  $e^+e^-\to\omega\pi^0$. In principle,
a similar picture should be observed in neutral channel
(\ref{eewpi}), as was explained above in the theoretical introduction. 
Really the whole situation here looks more complicated because of other
$\phi$ meson neutral decays like  $\phi\to\rho^0\pi^0$,
$\phi\to f_0\gamma,\,\sigma\gamma$ \cite{f0gamma}, which have
the same final state and interfere with the process (\ref{eewpi}).
The interference amplitude with the  $\phi\to\rho^0\pi^0$ decay is 
expected to be about 10\%, which is close to the value 17\% due to
$\phi\to\omega\pi^0$ decay, obtained in \cite{PhiWpi}.
In our preceding study \cite{EXP97} of the reaction (\ref{eewpi}),
we did not observe the interference because of small statistics 
and nonresonant background.

  In the present work the analysis given above also does not reveal
the interference effect with certainty in spite of higher statistics, 
because the background is still high. Here we present another analysis
specially dedicated to the interference observation.

The events selected by the criteria, described above, were
divided into following 3 classes:
$$
\begin{array}{llll}
1) & \chi^2_{\pi^0\pi^0\gamma} <20,    & |M_{\pi^0\gamma}-782|<30,  
& \zeta_\gamma < -5; \\
2) & \chi^2_{\pi^0\pi^0\gamma} <20,    & 30<|M_{\pi^0\gamma}-782|<60,  
& \zeta_\gamma < -5; \\ 
3) & \chi^2_{\pi^0\pi^0\gamma} <20,    & |M_{\pi^0\gamma}-782|<30,   
& N_\gamma=6.
\end{array}
$$
   The 6-photon events were put into the last class. It was done to
investigate a probability that sixth false photon
is superimposed on the  five photon event due to beam-related background. 
In the kinematic fit one of  photons was supposed to be spare.  

The simulation shows that the signal to background ratio
is maximal in  class 1, which was used later for investigation
of the interference effects. The contribution of 
process (\ref{eewpi})  in class 2 is about 10 times lower, which allows
to extract from the data the resonant background from the above mentioned
processes with $f^0\gamma\;,K_SK_L$ and $\eta\gamma$ intermediate states.
Using the ratio of the
resonant background in classes 1 and 2 obtained from the simulation,
one can estimate the background in class 1, when the resonant background 
in class 2 is determined from the experimental data.
The ratio of the 
background event numbers in the regions $|M_{\pi^0\gamma}-782|<30$ and
$30<|M_{\pi^0\gamma}-782|<60$ weakly depends on the limits on 
$\chi^2_{\pi^0\pi^0\gamma}$ and $\zeta_\gamma$ parameters.
For the process (\ref{phif0g}) this ratio varies in the limits
$1.09\div1.12$, while for the process (\ref{phietg})  --- from 0.9
to 1.07. These variations are within the statistical accuracy of the
corresponding MC simulation.
The value $1.1\pm 0.2$ was taken for this ratio.

 The visible cross section in each class was presented in the
following form:       
$$\sigma_{vis} = \alpha_i\epsilon [1+\delta(s)]\sigma(s)+
\beta_i\sigma_B(s)$$
\noindent where $\sigma(s)$ is given by (\ref{sgfit}),
$\varepsilon$ is the total over all classes detection efficiency
for the $e^+e^-\to\omega\pi^0\to\pi^0\pi^0\gamma$
process, and $\delta(s)$ represents the standard radiative corrections
\cite{RC}. As above, $\sigma_B(s)$ is assumed to coincide with
$\sigma(e^+e^-\to\eta\gamma\to 3\pi^0\gamma )$.
The parameters
$\alpha_i$ are the probabilities for our process
to be found in i-th class, the parameters  $\beta_i$ represent
nonresonant cross section in i-th class, normalized to the 
$e^+e^-\to\phi\to\eta\gamma$ process cross section. As it was already 
mentioned, the $\phi$ meson
excitation curve was described by the main background 
process (\ref{phietg}). The differences in excitation curves
for different background processes
were neglected at the present level of accuracy. The fitting
of the visible cross section was done for all three classes simultaneously.
The parameters
$\sigma_0$, $\sigma^\prime$, $Re(Z)$, $Im(Z)$, $\alpha_i$, $\beta_i$ except
$\beta_1$ and $\alpha_1$ were free. $\beta_1$ were found from
the expression $\beta_1=(1.1\pm 0.2)\cdot\beta_2$,
$\alpha_1$ -- from normalization $\sum\alpha_i=1$.
The detection efficiency $\varepsilon$=$27.5\%$,
obtained from the simulation, does not depend on the energy.
The coefficients  $\alpha_i$ are also energy independent.
The total number of fit parameters is 8. In each class the cross
section was measured in 15 points. The following values of interference
parameters were obtained 
\begin{eqnarray}
&Re(Z)=0.036\pm0.052,&\nonumber\\
&Im(Z)=-0.186\pm0.063.& \label{vdres1}
\end{eqnarray}
The statistical errors for this parameters are much higher than  the 
systematical ones. The quoted errors include
the systematics due to background subtraction in class 1.  

The ratio $\alpha_3/(\alpha_1+\alpha_3)$ gives the probability that the 
signal event
will be lost because the additional background photon is superimposed on it.
This probability turned out to be $0.04\pm 0.02$ and the corresponding
correction factor was included in the $\sigma_0$ and $\sigma^\prime$
determination in the previous section.

\begin{figure}[htb]
\centerline{\epsfig{figure=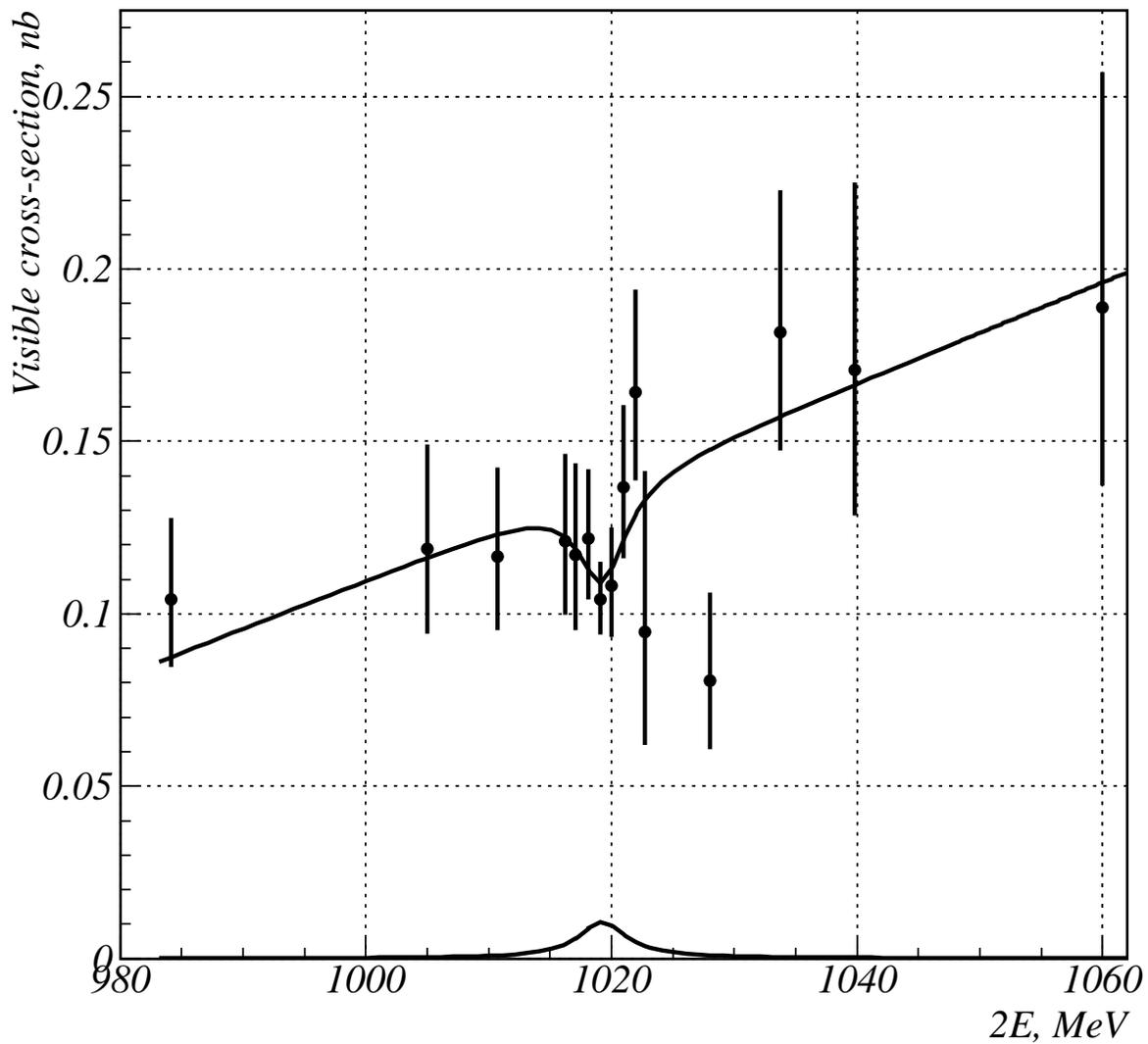, 
                               ,  height=14.5cm}}
\caption{Detected cross-section for the process
$e^+e^- \to \omega \pi^0 \to \pi^0 \pi^0 \gamma $ 
under cuts of group 1 and optimal fit.
The fitted resonant background is also shown at bottom.}
\label{Fig7}
\end{figure}

The detected cross section for class 1 and the fitted curve with
$\chi^2/d.f.=11.9/11$ are shown in Fig.\ref{Fig7}. The fitted
resonant background is also shown at the bottom. One could see,
that in spite of imposed strong cuts  in class 1, the resonant background
is about one third of interference amplitude wave and is the dominant
source of systematic error in  $Z$. This systematic error is estimated to
be about 6\%.

\section{Conclusions.}
In conclusion, we obtained in the present
work the following values of the $e^+e^-\to\omega\pi^0\to\pi^0\pi^0\gamma$
process cross section parameters: 
\begin{eqnarray}
&\sigma_0=(0.64\pm 0.08 )~{\mathrm nb},&\nonumber \\
&\sigma^\prime=(0.0048 \pm 0.0027 )~{\mathrm nb/MeV},&\nonumber \\
&Re(Z)=0.036 \pm 0.052,&\nonumber\\
&Im(Z)=-0.186 \pm 0.063.&\label{finres}
\end{eqnarray}

The measured 
interference amplitude is three standard deviation above zero. The measured 
in  \cite{PhiWpi} interference amplitude for the decay  
$\phi\to\omega\pi^0$ is
$Re(Z)=0.104\pm 0.029$, $Im(Z)=-0.118\pm 0.031$. Our calculation of the
interference amplitude for the decay $\phi\to\rho\pi^0\to\pi^0\pi^0\gamma$,
in the framework of the model described in the theoretical introduction,
gives $Re(Z)=-0.079$, $Im(Z)=-0.053$. The sum of these contributions
$Re(Z)=0.025$, $Im(Z)=-0.171$ agrees with our measurement. Otherwise
stated, if one subtract from the interference amplitude, measured
in this work, the expected contribution from the 
$\phi\to\rho\pi^0$ transition
given above, the $\phi\to\omega\pi$ decay branching ratio can be estimated
to be $5.4\cdot 10^{-5}$ with almost 100\% errors, which should be
compared to $(4.8^{+1.9}_{-1.7}\pm 0.8)\cdot 10^{-5}$ from \cite{PhiWpi}.
Note that we did  not make a theoretical estimate for contribution into 
interference amplitude from the processes
$\phi\to f_0\gamma,~\sigma\gamma\to\pi^0\pi^0\gamma$. More experimental
information about these transitions is required to estimate such 
contribution correctly.

The measured nonresonant cross section  at $\phi$-meson
$\sigma(e^+e^-\to\omega\pi^0)=
\sigma(e^+e^-\to\omega\pi^0\to\pi^0\pi^0\gamma)/Br(\omega\to\pi^0\gamma)=
(7.5\pm 0.9)~{\mathrm nb}$ agrees with the result 
$(8.7\pm 1.0\pm 0.7)~{\mathrm nb}$
from  \cite{ND}
and with the result $(8.2\pm 0.2\pm 0.9)~{\mathrm nb}$ from \cite{PhiWpi} 
in channel with 
charged pions $e^+e^-\to\omega\pi^0\to\pi^+\pi^-\pi^0\pi^0$, as well as
with the preliminary CMD-2 result \cite{CMD2} 
$(7.2\pm 0.5)~{\mathrm nb}$.
Such nonresonant cross section can not be explained by only $\rho$-meson
contribution, which is about 1.6 times lower. Although the inclusion of the
$\rho^\prime(1465)$ meson in the fit improves somewhat the agreement 
between the theoretical prediction and the experiment, the question about 
$\rho$-meson radial excitations lies beyond the scope of this work, because 
it requires an experimental information for higher energies.

At last, in Table 4 we provide numerical values of the measured cross
section $\sigma(e^+e^-\to\omega\pi^0\to\pi^0\pi^0\gamma)=\frac{\sigma_v(s)}
{\epsilon [1+\delta(s)]}$.

\begin{table}[htb]
\begin{center}
\begin{tabular}{|c|c|}
\hline
 $\sqrt{s}, \; \;$ MeV & cross section, nb \\
\hline
 984.1   &  0.57  $\pm$   0.11 \\    
 1005.0  &  0.56  $\pm$   0.14 \\    
 1010.7  &  0.47  $\pm$   0.10 \\    
 1016.2  &  0.55  $\pm$   0.10 \\
 1017.1  &  0.48  $\pm$   0.10 \\    
 1018.1  &  0.56  $\pm$   0.08 \\
 1019.1  &  0.59  $\pm$   0.05 \\
 1020.0  &  0.51  $\pm$   0.07 \\
 1021.0  &  0.74  $\pm$   0.11 \\    
 1021.9  &  0.81  $\pm$   0.14 \\    
 1022.7  &  0.55  $\pm$   0.21 \\    
 1028.0  &  0.53  $\pm$   0.12 \\    
 1033.7  &  0.97  $\pm$   0.19 \\    
 1039.8  &  0.96  $\pm$   0.25 \\    
 1060.0  &  1.08  $\pm$   0.31 \\ 
\hline
\end{tabular}
\end{center}
\caption
{Measured cross section $\sigma(e^- e^+ \to \omega \pi^0 \to \pi^0 \pi^0
\gamma)$ (mean values between Cut I and Cut II, corrected by the factor 1.04,
as discussed in the text). Only statistical errors are indicated. Systematic
errors are estimated to be about 5\%.}
\label{tb4}
\end{table}

\section{Acknowledgement}
This work is supported in part by Russian Fund for basic researches, grants
No. 99-02-16813 and 96-15-96327.

\newpage
\end{document}